\documentclass[twocolumn,showpacs,preprintnumbers,amsmath,amssymb,a4paper]{revtex4}

\usepackage{amsmath,amssymb,bbm,epsfig}

\newcommand{\be}{\begin{equation}}
\newcommand{\ee}{\end{equation}}
\newcommand{\ba}{\begin{array}}
\newcommand{\ea}{\end{array}}
\newcommand{\bqa}{\begin{eqnarray}}
\newcommand{\eqa}{\end{eqnarray}}

\newcommand{\tr}{\mbox{Tr}}
\newcommand{\bra}[1]{\ensuremath{\langle #1 |}}
\newcommand{\ket}[1]{\ensuremath{| #1 \rangle}}

\begin{document}

\title{Entanglement measures as physical observables}

\author{Florian Mintert}
\affiliation{
Department of Physics, Harvard University,
17 Oxford Street, Cambridge Massachusetts, USA}

\date{\today}

\begin{abstract}
We discuss why regular observables can not be proper entanglement measures,
and how observables in a generalized setting can be used to make an entanglement monotone
a directly observable quantity for the case of pure states.
For the case of mixed states, these generalized observables can be used to find valid lower bounds
on entanglement monotones that can be measured in laboratory experiments in the same fashion
as it can also be done for pure states.
\end{abstract}


\maketitle

Quantum entanglement constitutes a fundamental qualitative difference of many body quantum systems
as compared to their classical counterparts,
and its signatures are observed in various ways.
For example the success of a teleportation protocol \cite{teleport}is an unambiguous evidence of entanglement.
And, there are also more direct ways to experimentally verify the existence of entanglement in a
quantum system, such as entanglement witnesses \cite{witnesshor,witness04}, or Bell inequalities
\cite{epr,bell,costas,PhysRevLett.49.91,PhysRevLett.49.1804}.
Those quantities provide reliable tools for the verification of entanglement,
and they have been used to prove the successful preparation of entangled states
in various laboratory experiments \cite{W8,GHZ6}.

But, although signatures of entanglement can be observed in actual laboratory experiments,
entanglement {\em per se} is not a regular physical observable.
That is, there is no hermitean operator such that the value of an entanglement measure
could be given by its expectation value for {\em any} state of a composite quantum system.

On the other hand, the experimental quantification of entanglement is a vital ingredient both for
fundamental investigations, such as tests of the emergent classical behavior of large quantum systems,
as well as for more application oriented tasks, as for example the establishment of entanglement
over large distances for purposes of secure communication \cite{gisin50km,grancanaria}.
Typically, an entanglement measure can {\em not} be measured, but rather has to be evaluated
what requires the knowledge of the complete density matrix.
Left aside the mathematical issues for such an evaluation, the experimental reconstruction of the density matrix,
requires the determination of a complete set of observables;
and since the number of such observables grows rapidly with the dimension of the system,
such a {\it quantum state tomography} \cite{RevModPhys.29.74} is a viable solution only for rather small systems.

Any entanglement measure satisfies a rather strong invariance condition --
that is, it has to be invariant under arbitrary local unitary transformations.
Regular observables, however, typically do not have this invariance,
and as a consequence thereof, one has to measure various such observables
in order to construct an invariant quantity.
In the sequel we will elaborate on how invariant observables can be found
in a generalized setting of measurements performed on quantum systems,
and how entanglement monotones \cite{vidal00} can be expressed
in terms of a only few of such observables.

\section{Measuring non-linear functionals}

There is an abundance of proposed inequivalent entanglement monotones, or -measures for quantum states
of composite systems, {\it i.e.} a systems that decompose into two or more subsystems.
Formally, the term `composite' implies the the Hilbert space ${\cal H}$ that describes the system
is given by the tensor product of the subspaces ${\cal H}_i$ that describe the individual subsystems.

Virtually the only property that all those inequivalent quantities share is that all of them
are non-linear functionals of the quantum state whose entanglement properties they aim to describe.
A common example is the v. Neumann entropy $S=-\tr\varrho_r\ln\varrho_r$ of the reduced density matrix $\varrho_r$,
which is obtained by tracing over one subsystem of a pure bipartite quantum state $\ket{\Psi}\bra{\Psi}$.

Therefore, a necessary requirement for a direct observation of such an entanglement measure is the ability
to measure non-linear properties of quantum states.
Such a task is indeed possible after the repeated preparation of the same state in an identical fashion
\cite{pereswootters,ekert:217901,toddbrun,quantph0604109,quant-ph/0606017}.
Typically, in an experiment with single quantum systems, a state is prepared repeatedly,
and after each preparation a measurement is performed, so that after many repetitions reliable measurement statistics is obtained.
If one, however, waits with the measurement after several repetitions of the preparation,
then one is able to perform a measurement on
$n$ identically prepared quantum systems, instead of a single one.
Doing so, one can measure collective observables on $n$ `copies'
$\tr A_n\varrho^{\otimes n}$ instead of $\tr A_1\varrho$.

If the experimental preparation works like a perfect source that repeatedly emits the same state $\ket{\Psi}$,
then one has indeed an $n$-fold copy $\ket{\Psi}^{\otimes n}$ after $n$ repetitions.
Though, what happens, if the preparation is imperfect, and one has to treat it like an imperfect source
that emits different states $\ket{\Psi_i}$ with corresponding probabilities $p_i$, what gives rise to a density matrix
$\varrho=\sum_ip_i\ket{\Psi_i}\bra{\Psi_i}$?
At each emission the source produces the state $\ket{\Psi_i}$ with probability $p_i$.
Thus, after $n$ emissions the string
$\ket{\Psi_{i_1}}\otimes\ket{\Psi_{i_2}}\otimes\hdots\otimes\ket{\Psi_{i_n}}$ has been prepared with
probability $p_{i_1}p_{i_2}\hdots p_{i_n}$.
After $N\gg 1$ repetition of this $n$-fold flawed emission,
this is equivalent to the repeated flawless preparation of an $n$-fold copy of the density matrix $\varrho$: 
\be
\sum_{i_1,\hdots,i_n}p_{i_1}\hdots p_{i_n}
\ket{\Psi_{i_1}}\bra{\Psi_{i_1}}\otimes\hdots\otimes\ket{\Psi_{i_n}}\bra{\Psi_{i_n}}=\varrho^{\otimes n}\ .
\ee
Thus, given this ability to produce identically prepared quantum states,
one can observe experimentally collective properties, such as
$\tr A_2\varrho\otimes\varrho$, $\tr A_3\varrho\otimes\varrho\otimes\varrho$, or
$\tr A_n\varrho^{\otimes n}$.

\section{Invariant Observables}

A necessary requirement for a quantity to be an entanglement monotone is that it is invariant under
arbitrary local unitary transformations.
Since this has to be satisfied for $\tr\varrho^{\otimes n}A$, for {\em any} state $\varrho$,
we require that the operator $A$ itself is invariant under local unitaries.
That is, we need to find invariant observables $A_n$ in an $n$-fold Hilbert space ${\cal H}^{\otimes n}$:
\be
A=u_1^{\otimes n}\otimes u_2^{\otimes n}\ A\ \bigl(u_1^\dagger\bigr)^{\otimes n}\otimes \bigl(u_2^\dagger\bigr)^{\otimes n}\ ,
\ee
for arbitrary unitaries $u_1$ (acting on ${\cal H}_1$), and $u_2$ (acting on ${\cal H}_2$),
where we have been assuming the bipartite case ${\cal H}={\cal H}_1\otimes{\cal H}_2$ for simplicity.
The following reasoning, however, also generalizes to the multipartite case in a straight forward manner.

Like any operator, one can expand $A$ in terms of a complete set of {\em local} operators
 $A=\sum_{ij}A_{ij}\sigma_i\otimes\kappa_j$,
 where $\{\sigma_i\}$ is a complete set of operators acting on ${\cal H}_1^{\otimes n}$,
 and  $\{\kappa_j\}$ is a complete set of operators acting on ${\cal H}_2^{\otimes n}$.
 Since $A$ is invariant under local unitaries,
 it also equals its avarage over all such transformations
 \bqa
 A&=&\int d\mu(u_1)d\mu(u_2)\ u_1^{\otimes n}\otimes u_2^{\otimes n}\ A\
 \bigl(u_1^\dagger\bigr)^{\otimes n}\otimes \bigl(u_2^\dagger\bigr)^{\otimes n}\nonumber\\
 &=&\sum_{ij}A_{ij}\tilde \sigma_i\otimes\tilde\kappa_j\ ,
 \eqa
with
\bqa
\tilde\sigma_i&=&\int d\mu(u_1)\ u_1^{\otimes n}\sigma_i\bigl(u_1^\dagger\bigr)^{\otimes n}\ ,\\
\tilde\kappa_j&=&\int d\mu(u_2)\ u_2^{\otimes n}\kappa_i\bigl(u_2^\dagger\bigr)^{\otimes n}\ .
\eqa
The averaged operators $\tilde\sigma_i$, and $\tilde\kappa_j$
now are local operators that are invariant under arbitrary local unitary transformations.
That is, any global operator that is invariant under local unitaries can always be decomposed into
local invariant operators.
This does not only allow us to restrict the search for invariant observalbes to local ones --
what is a rather technical advantage --
but it also implies that any measurement of an invariant observable can always be decomposed
into local measurements without loosing the advantage of measuring invariant observables.

Therefore, we will be looking for invariants in an $n$-fold Hilbert space $h^{\otimes n}$,
where $h$ can be any of the subspaces ${\cal H}_i$ associated with a composite quantum system.
Let us start out with an observable on a single Hilbert space $h$.
Since $A$ needs to be invariant under arbitrary unitaries $u$,
it -- in particular -- has to be invariant under infinitesimal transformations ${\mathbbm 1}+i\varepsilon h$,
where $h$ is hermitean.
For such infinitesimal transformations the invariance condition reduces to $\left[A,h\right]=0$.
That is, we need an observable that commutes with arbitrary hermitean operators,
and this condition is satisfied only for the identity ${\mathbbm 1}$.
Indeed, the fact that the identity is the only operator that is invariant under arbitrary
unitaries is well known from {\it Shur's lemma}.
If, however, we consider the case of multiple copies, an observable $A$
only needs to be invariant under unitary transformations of the form
$u^{\otimes n}$, what is a significantly smaller class than arbitrary unitaries in $h^{\otimes n}$.
And, indeed one can find non-trivial observables that satisfy this invariance.

Such invariant operators can be constructed in a particularly simple fashion with the help of permutation operators $\Pi$.
In the simplest case of a two-fold copy, there is the permutation operator
$\Pi_{12}$ that exchanges the two copies;
{\it i.e.} its action on a state $\ket{\Psi}\otimes\ket{\Phi}\in h\otimes h$ reads:
$\Pi_{12}\ket{\Psi}\otimes\ket{\Phi}=\ket{\Phi}\otimes\ket{\Psi}$.
For a three-fold copy there are permutations that exchange the states of two of the Hilbert spaces and leave the state of the third space unchanged,
such as $\Pi_{12}\ket{\Psi}\otimes\ket{\Phi}\otimes\ket{\Xi}=\ket{\Phi}\otimes\ket{\Psi}\otimes\ket{\Xi}$,
or the cyclic permutation with the action
$\Pi_c\ket{\Psi}\otimes\ket{\Phi}\otimes\ket{\Xi}=\ket{\Xi}\otimes\ket{\Psi}\otimes\ket{\Phi}$.
And, similarly, permutation operators can be found for any multiple product $h^{\otimes n}$ of a Hilbert space with itself.
Now, any permutation $\Pi$ on $h^{\otimes n}$ commutes with arbitrary $n$-fold unitaries $u^{\otimes n}$;
and since permutations are also hermitean, we have indeed found a set of invariant observables.
In the following, however, we will not focus on the permutation operators themselves,
but rather on the projectors onto their eigenspaces.
Since, however, any such projector $P$ can be written as a sum over powers of projectors,
also these projectors are valid invariant observables.

For the sake of specificity, let us look more closely into some exemplary case of such projectors,
and take $\Pi_{12}$ for a twofold product of $h$.
Two consecutive permutations $\Pi_{12}$ are the identity operation $\Pi_{12}^2={\mathbbm 1}$.
Therefore, $\Pi_{12}$ has eigenvalues $\lambda_\pm=\pm 1$.
The projector onto the space that is spanned by the eigenvectors that are
associated with the eigenvalue $\lambda_+=1$,
reads $P_+=(\Pi_{12}+\Pi_{12}^2)/2$,
and the projector onto the second eigenspace of $\Pi_{12}$ is
$P_-=(-\Pi_{12}+\Pi_{12}^2)/2$.
Now, $P_+$ is the projector onto the symmetric states
$\ket{\phi_i}\otimes\ket{\phi_j}+\ket{\phi_j}\otimes\ket{\phi_i}$,
that is those states that are invariant under the application of $\Pi_{12}$,
whereas the antisymmetric states
$\ket{\phi_i}\otimes\ket{\phi_j}-\ket{\phi_j}\otimes\ket{\phi_i}$,
that span the range of $P_-$, obtain a prefactor of `$-1$' upon $\Pi_{12}$.
In the common case of two-level systems,
there is a single anti-symmetric state -- the singlet
$(\ket{01}-\ket{10})/\sqrt{2}$, so the the projector $P_-$ onto the antisymmetric subspace is one-dimensional;
the projector $P_+$ onto the symmetric part in turn is three-dimensional,
and is comprised of the triplet states.

\section{Entanglement measures for pure states}

\begin{figure}
\begin{center}
\includegraphics[width=0.5\textwidth]{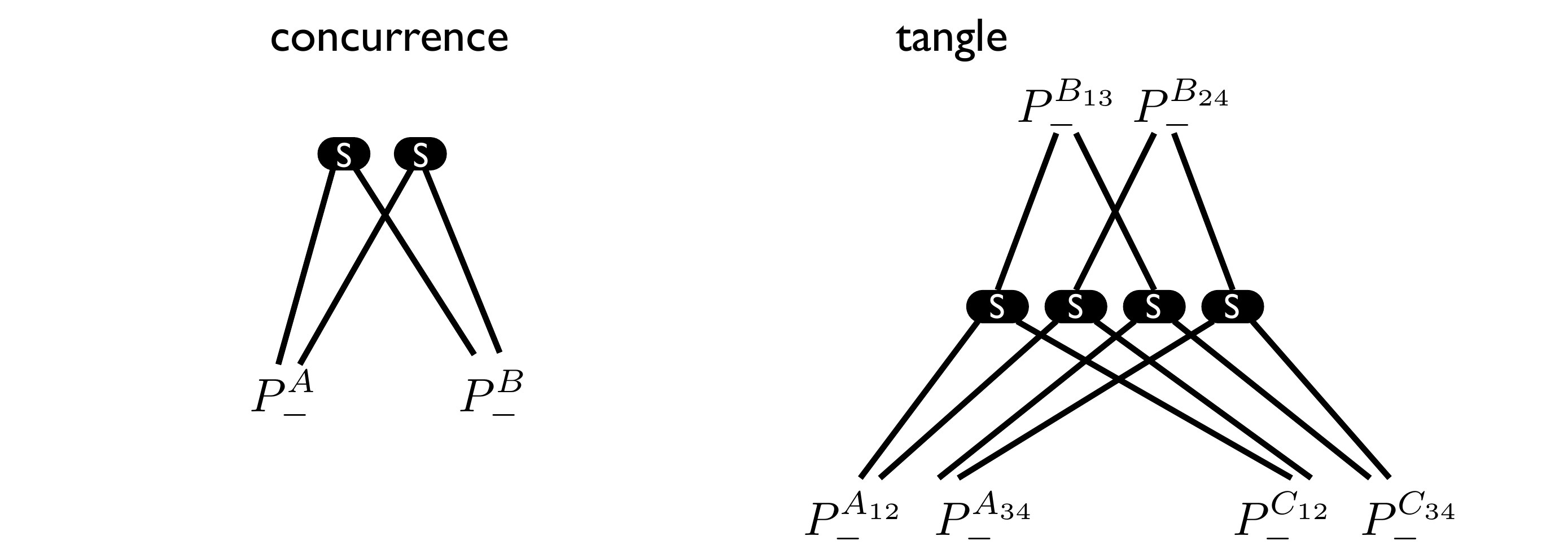}
\caption{
Schematic setting of the observables that yield concurrence (left) of a bipartite system,
and the tangle (right) of a three-partite system.
Each of the two equivalent sources $S$ emits a bipartite state $\ket{\Psi}$,
and each of the individual subsystems is symbolized by a black line.
The concurrence of the state $\ket{\Psi}$ is given in terms of the probability
to find both the two first-subsystem components and the two second-subsystem components
of the twofold state $\ket{\Psi}\otimes\ket{\Psi}$ in an antisymmetric state.
The tangle of a tri-partite system can be given in a similar fashion with
four identically prepared quantum systems.}
\label{fig:1}
\end{center}
\end{figure}

Now, one can indeed recover various well known entanglement measures in terms of the above invariant observables.
The most frequently used of those is probaly {\em concurrence} \cite{eof}.
It is obtained with the help of a twofold copy of a state $\ket{\Psi}$, and reads
\be
c(\Psi)=\sqrt{\bra{\Psi}^{\otimes 2}P_-^{A}\otimes P_-^{B}\ket{\Psi}^{\otimes 2}}\ ,
\label{concurrence}
\ee
where $A$ labels the first-, and $B$ the second subsystem.
That is $P_-^{A}$ acts on the first-subsystem components of both copies of $\ket{\Psi}$,
whereas $P_-^{B}$ acts on the two second-subsystem components
as shown schematically in Fig.~\ref{fig:1}.
In more practical terms Eq.~(\ref{concurrence}) states that the concurrence of an arbitrary pure state
is given in terms of the probability to finding both the first-subsystem
components and the second-subsystem components in an antisymmetric state.

Also the well known tangle \cite{tangle} for tripartite systems can be expressed in the present framework,
if observables on fourfold copies are invoked.
But although four copies are required, it is indeed sufficient to perform collective measurements only on pairs of
subsystems:
\be
\tau(\Psi)=\left(\bra{\Psi}^{\otimes 4}P^{A_{12}}_{-}P^{A_{34}}_{-}P^{B_{13}}_{-}P^{B_{24}}_{-}P^{C_{12}}_{-}P^{C_{34}}_{-}\ket{\Psi}^{\otimes 4}\right)^{\frac{1}{4}}\ .
\ee
Here, we dropped the symbol of the tensor product for brevity,
and the indices on $A$, $B$, and $C$ specify on which copies the corresponding operators are associated;
$P^{A_{12}}_{-}$, for example acts on copy `one' and `two' of subsystem $A$,
as sketched in Fig.~\ref{fig:1}.

\begin{figure}
\begin{center}
\includegraphics[width=0.5\textwidth]{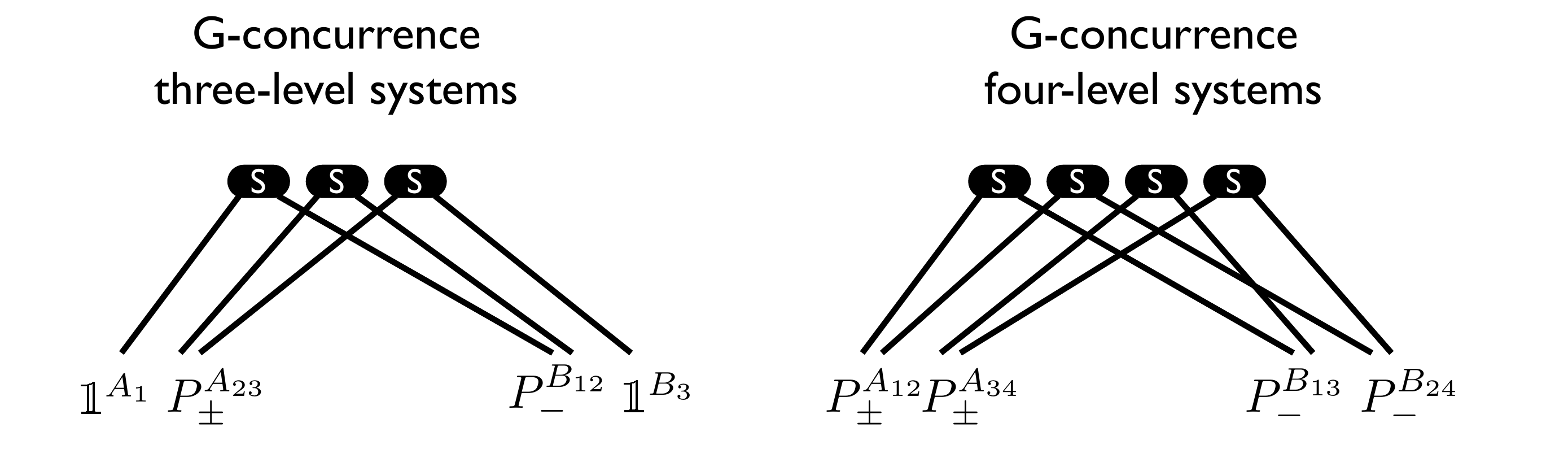}
\caption{
The G-concurrence can be expressed in terms of the present invariant observables,
as shown schematically similarly to Fig. \ref{fig:1}
For the case of three-level systems, it is given in terms of the probability
to observe the first-subsystem components of the first and second copy in an antisymetric state,
and the second subsystem components of the second and third subsystem in an antisymmetric state.
For a bipartite four-level system it is given with the help of a fourfold copy as depicted schematically.}
\label{fig:2}
\end{center}
\end{figure}

Tangle, and concurrence are -- although defined here for arbitrary dimensional systems -- typically used to characterize
two-level systems.
In particular, for a bipartite two-level system, the entanglement of pure states is characterized completely by concurrence,
and all other measures can be expressed as function of concurrence only.
For higher dimensional systems, however, this is no longer true, and more than a single monotone is required for an
exhaustive characterization of the entanglement properties, even of pure states.
Just to give a few examples that allow to see, how various entanglement monotones can be expressed in terms of the present invariant observables, we show how the so-called $G$-concurrence $c_G$ \cite{gour:012318} is readily expressed
within the present framework.
In terms of the Schmidt coefficients $\{\lambda_i\}$ of a state $\ket{\Psi}$ of a $d\times d$ system,
 the $G$-concurrence reads $c_G=(\Pi_{i=1}^{d}\lambda_i)^{\frac{1}{d}}$.
 That is, $c_G$ vanishes for all states that are not of maximal Schmidt rank.
 For the case of a $3\times 3$-system $c_G$ reads
\be
c_G(\Psi)=\left(\bra{\Psi}^{\otimes 3}
{\mathbbm 1^{A_{1}}\left(P_{-}^{A_{23}}-\frac{P_{+}^{A_{23}}}{3}\right)P_{-}^{B_{12}}{\mathbbm 1}^{B_{3}}}
\ket{\Psi}^{\otimes 3}\right)^{\frac{1}{3}}\ ,
\ee
in terms of the projectors $P_{\pm}$ acting on the respective subsystem components of two of the three copies
of $\ket{\Psi}$.
${\mathbbm 1}$ is the identity, and it implies that no measurement has to be performed on the corresponding subsystem.
Analogously, also for a $4\times 4$ system the $G$-concurrence can be found in terms of the invariant observable
\be
A=(P_{-}^{A_{12}}P_{-}^{A_{34}}-\frac{P_{+}^{A_{12}}P_{+}^{A_{34}}}{3})P_{-}^{B_{13}}P_{-}^{C_{24}}
\ee
on a fourfold copy of $\ket{\Psi}$,
as also shown schematically in Fig.~\ref{fig:2}.

\section{Entanglement measures for mixed states}

All the above considerations concerning invariant observables directly apply also to mixed states.
However, none of the above quantities are valid entanglement monotones for mixed states,
but only for pure ones;
and a major difference between entanglement monotones for pure states,
and those of mixed states is that in the latter case typically no closed form of the functional dependence
on the state is known.
The concurrence for $2\times 2$ systems \cite{wootters98}, and the negativity \cite{vidal02} are among the few exceptions, 
where the functional dependence is known at least implicitly,
and both quantities can be measured exactly on up to four identically prepared quantum states
in the $2\times 2$ case \cite{phorodecki03,carteret:040502,quantph0604109}.
The negativity, and other separability conditions can also be measured for higher dimensional systems \cite{horodecki:052323}.
However, apart from these special cases, the direct measurement of proper entanglement monotones is
not possible because no entanglement monotone can be given as functional of a quantum state in closed form,
but typically it is given in terms of a mathematical optimization problem.
Also, as the studies of \cite{horodecki:052323} suggests, an exact characterization of an entanglement monotone
is expected to require measurements on a number of copies that increases with the dimension of the system
to be analyzed.
The necessity to provide simultaneously a too large number of identically prepared quantum systems,
however, makes such an approach infeasible for large quantum systems.
Therefore, we present here a compromise of accuracy and effort.
That is we will satisfy ourselves with lower bounds on entanglement monotones,
but restrict the number of simultaneous copies on which measurement are to be performed.
A lower bound is a reliable quantity, and the restriction to a maximum number of copies guarantees
suitable scaling behavior.

Any entanglement monotone ${\cal M}$ for pure states can be generalized to mixed states via the
{\it convex roof} construction
\be
{\cal M}(\varrho)=\inf_{\{ p_i,\ket{\Psi_i}\} }\sum_i\ p_i{\cal M}(\Psi_i)\ ,
\ee
where the infimum is taken over all decompositions $\varrho=\sum_ip_i\ket{\Psi_i}\bra{\Psi_i}$
of $\varrho$ into pure states.
All the pure state monotones that we have been discussing above are homogeneous functions in the density matrix,
{\it i.e.} they satisfy ${\cal M}(p\varrho)=p{\cal M}(\varrho)$.
This allows to reformulate the convex roof construction in terms of {\it subnormalized} states
$\ket{\psi_i}=\sqrt{p_i}\ket{\Psi_i}$,
so that the infimum is to be taken among all decompositions $\varrho=\sum_i\ket{\psi_i}\bra{\psi}$
into subnormalized states.
Given that, we can now describe a general prescription of how to find bounds on convex roof
entanglement monotones whose pure state counterpart is of the form
\be
{\cal M}(\Psi)=\left(\bra{\Psi}^{\otimes n}A_n\ket{\Psi}^{\otimes n}\right)^{\frac{1}{n}}\ .
\ee
What one has to find is an operator $V_n$ -- that, in turn is constructed to be an invariant observable --
such that the inequality
\be
\prod_{i=1}^{n}c(\psi_i)\ge
\left(\bigotimes_{i=1}^{n} \bra{\psi_i}\right)V_n
\left(\bigotimes_{i=1}^{n} \ket{\psi_i}\right)
\label{inequality}
\ee
holds for arbitrary states $\{\ket{\psi_i}\}$.
Although, there is no general prescription of how to prove such a relation,
such an algebraic inequality is significantly easier to handle than the original optimization problem
defined by the convex roof construction above.
Given Eq.~(\ref{inequality}), one can then find the desired bound on ${\cal M}(\varrho)$
for arbitrary mixed states in the following fashion:
\bqa
\left({\cal M}(\varrho)\right)^n&=&\inf\sum_{i_1i_2\hdots i_n}\prod_{j=i_1}^{i_n}{\cal M}(\psi_j)\label{eq:1}\\
&\ge&\sum_{i_1i_2\hdots i_n}\bigl(\bigotimes_{j=i_1}^{i_n}\bra{\psi_j}\bigr)V_n\bigl(\bigotimes_{j=i_1}^{i_n}\ket{\psi_j}\bigr)\label{eq:2}\\
&=&\tr\ \varrho^{\otimes n}\ V_n\ .\label{eq:3}
\eqa
Here, we arrived at Eq.~(\ref{eq:2}), staring out with Eq.~(\ref{eq:1}), and making use of Eq.~(\ref{inequality}).
Since the right hand side of Eq.~(\ref{eq:2}) does not depend on in which pure state decomposition of $\varrho$
it is evaluated, the infimum can be dropped there.
That is, we end up with a general lower bound on a given monotone ${\cal M}$ for arbitrary mixed states
that can be measured on $n$ identically prepared quantum states thereof.

\subsection{Storage errors}

So far, we have been assuming that all the $n$ copies on which a measurement is performed are really identically prepared.
However, in actual laboratory experiments, such a condition will typically not be given with perfection.
Since all the $n$ copies have to be available at the same time, they either have to be prepared in parallel
using different sources,
or they are prepared sequentially by the same source, and stored until all $n$
quantum states have been prepared.
In both cases, however, experimental imperfections might lead to the preparation of an $n$-fold string of
different states $\varrho_1\otimes\varrho_2\otimes\hdots\otimes\varrho_n$,
rather than $\varrho\otimes\varrho\otimes\hdots\varrho$.
In that case, exactly the same reasoning as above in Eqs.~(\ref{eq:1})-(\ref{eq:3}) leads to the conclusion
that the geometric mean of ${\cal M}(\varrho_i)$ is bounded from below by the expectation values of $V_n$:
\be
\prod_{i=1}^{n}{\cal M}(\varrho_i)\ge\tr\ \bigotimes_{i=1}^{n}\varrho_i\ V_n\ .
\ee
In the case of sequential preparation those states that have been prepared earlier have to be stored for a longer time
than those that are prepared at a later time, so that they will have more time to decohere.
In the typical situation that entanglement is to be established over a macroscopic distance,
each of the subsystems will interact with its local environment,
so that decoherence is a purely local effect that can not increase entanglement.
Therefore, the value of any entanglement monotone can only decrease during storage,
what allows to conclude that the expectation value of $V_n$
obtained with those imperfect `copies' also provides a lower bound on
${\cal M}$ for the state $\varrho$ before storage:
\be
\left({\cal M}(\varrho)\right)^n\ge \prod_{i=1}^{n}{\cal M}(\varrho_i)\ge \tr\ \bigotimes_{i=1}^{n}\varrho_i\ V_n\ .
\ee
Therefore, even under imperfect conditions, one still obtains a reliable outcome.

\subsection{Concurrence}

To get more specific, let us focus on the case of concurrence, and measurements of two identically prepared quantum states.
In this case there are four invariant observables
$P_-\otimes P_-$, $P_-\otimes P_+$, $P_+\otimes P_-$, and $P_+\otimes P_+$,
but only three of them are independent, since they sum up to unity.
The concurrence of pure states is given exactly in terms of $P_-\otimes P_-$ only.
By continuity one therefore expects that the expectation value of this operator
also gives a reasonable estimate for the concurrence of very weakly mixed states.
However, this expectation value can overestimate concurrence.
The expectation value of $P_-\otimes P_+$, or $P_+\otimes P_-$ with respect to a twofold state $\varrho\otimes\varrho$
on the other hand  vanishes exactly if $\varrho$ is {\em pure},
and indeed these two observables allow to estimate the mixing of a given state:
$\tr\ \varrho\otimes\varrho\ (P_-\otimes P_++P_+\otimes P_-)\sim(\tr\varrho)^2-\tr\varrho^2$.
Thus, one can get a -- so far still qualitative -- idea of by how far concurrence might be overestimated.
In order to find a lower bound on concurrence, one has to subtract some contribution from
$P_-\otimes P_-$ in order to rather underestimate concurrence;
and since this contribution should be connected with the mixing of the quantum state $\varrho$,
we use the ansatz
$V=P_-\otimes P_--\alpha_1P_-\otimes P_+-\alpha_2P_+\otimes P_-$,
where $\alpha_{1/2}$ are positive prefactors that have to be adjusted so that Eq.~(\ref{inequality})
be satisfied for this choice of $V$.
And, indeed such prefactors can be found, and one has a valid lower bound on the concurrence of
arbitrary mixed states for $1\ge\alpha_1\ge 0$, and $\alpha_2=1-\alpha_1$ \cite{mcmixed}.
In particular for weakly mixed states, this bound gives a very good estimate of the actual value of concurrence,
and also for states with substantial mixing the bound yields surprisingly good assessment \cite{mcmixed};
but one should not expect to be able to recover the exact border between separable and entangled states with such means,
since such an investigation requires more knowledge on a state then just the few expectation values
that are utilized here.

\section{Outlook}

The present approach allows for a systematic construction of observables to be measured ofnseveral identically prepared quantum states,
many of which have been proven to be entanglement monotones for pure states \cite{gour:012318,rafal}.
Lower bounds for all those monotones can be derived for mixed states, given knowledge of properties
of the respective observables on pure states only.
And, although currently only the case of measurements on two identically prepared quantum system
has been generalized to mixed states,
the significantly facilitated situation of pure states as compared to mixed ones,
gives reason to expect that also monotones invoking more than two copies can be generalized to mixed states.

The case of concurrence \cite{mcmixed} shows that the measurable bounds are the tighter the weaker a state is mixed,
and one might wonder wether one can improve the present bounds in order to obtain better results for highly mixed states.
Currently, this question can not be answered rigorously but it is expected that measurements on an increasing number
of copies will yield tighter bounds,
and it would be very nice to find observables $V_n$, such that the measurable bounds
converge to the exact value of concurrence, or an other monotone for measurements on an increasing number $n$
of identically prepared quantum states.

\section{Acknowledgment}

We would like to thank for the stimulating discussions and the intensive, and fruitful collaboration with 
Andreas Buchleitner, Rafa\l\ Demkowicz Dobrzanski, Marek Ku\'s, Leandro Aolita, Stephen Walborn
Paulo Souto Ribeiro, and Luiz Davidovich.
The corresponding financial support by VolkswagenStiftung
(under the project `Entanglement measures and the influence of noise'),
German Academic Exchange Service (DAAD) in terms of a PostDoc stipend and a joint DAAD/Capes
binational project,
and Alexander v. Humboldt foundation is gratefully acknowledged.

\bibliography{../../referenzen}

\end{document}